\documentclass[12pt]{article}

\usepackage[latin1]{inputenc}
\usepackage[T1]{fontenc}
\usepackage[english]{babel}
\usepackage{amsfonts}
\usepackage{amssymb}
\usepackage{amsmath}
\usepackage{amsthm}
\usepackage{graphicx}
\def\be{\begin{equation}}
\def\ee{\end{equation}}
\def\bc{\begin{center}}
\def\ec{\end{center}}
\makeindex

\newtheorem{proposition}{Proposition}

\newtheorem{theorem}{Theorem}

\newtheorem{remark}{Remark}
\newtheorem{definition}{Definition}

\def\be{\begin{equation}}
\def\ee{\end{equation}}
\def\bc{\begin{center}}
\def\ec{\end{center}}
\usepackage[latin1]{inputenc}
\usepackage[english]{babel}

\usepackage{color}
\usepackage{latexsym}
\usepackage{amssymb}
\usepackage{amsfonts}
\usepackage{amsmath}
\usepackage{graphicx}

\begin{document}

\title{Notes on ferromagnetic diluted P-spin model}
\author{Elena Agliari\footnote{Dipartimento di Fisica, Universit\'a di Parma and INFN, Gruppo Collegato di Parma}
\and Adriano Barra\footnote{Dipartimento di Fisica, Sapienza
Universit\'a di Roma} \and Federico Camboni\footnote{Dipartimento
di Fisica, Sapienza Universit\'a di Roma}}

%
%
%

\maketitle

\begin{abstract}
In this paper we develop the interpolating cavity field technique
for the mean field ferromagnetic p-spin. The model we introduce is
a natural extension of the diluted Curie-Weiss model to $p>2$ spin
interactions. Several properties of the free energy are analyzed
and, in particular, we show that it recovers the expressions
already known for $p=2$ models and for $p>2$ fully connected
models. Further, as the model lacks criticality, we present
extensive numerical simulations to evidence the presence of a
first order phase transition and deepen the behavior at the
transition line. Overall, a good agreement is obtained among
analytical results, numerics and previous works. \end{abstract}

\section{Introduction}
Born as a theoretical background for thermodynamics, statistical
mechanics provides nowadays a flexible approach to several
scientific problems whose depth and wideness increases
continuously. In fact, in the last decades statistical mechanics
has invaded fields as diverse as spin glasses \cite{MPV}, neural
networks \cite{amit}, protein folding \cite{huang}, immunological
memory \cite{immune}, social networks \cite{PL1}, theoretical
economy \cite{ton} and urban planning \cite{intro2}. As a
consequence, an always increasing need for models and proper
techniques must be fulfilled. Coherently, recently, several models
have been systematically tackled via the smooth cavity field by
the authors, namely the Curie-Weiss model \cite{abarra}, the fully
connected $p$-spin model \cite{abarra3}, the
Sherrington-Kirkpatrick model \cite{barra1}, its diluted counter
part Viana-Bray model \cite{bdsf} and the diluted ferromagnetic
model \cite{ABC}. All these models can just be seen as different
components of a more general class including models based on
binary agents with mean field interactions (Fig. \ref{fig:cubo}).
Now, in order to complete the analysis of the free energies for
the whole class, the X-OR-SAT (of the Random Optimization Theory
\cite{xorsat}) and the diluted ferromagnetic $p$-spin model, are
still missing; this paper is devoted to the study of the latter.

\begin{figure}[tb]\label{fig:cubo}
\includegraphics[height=100mm]{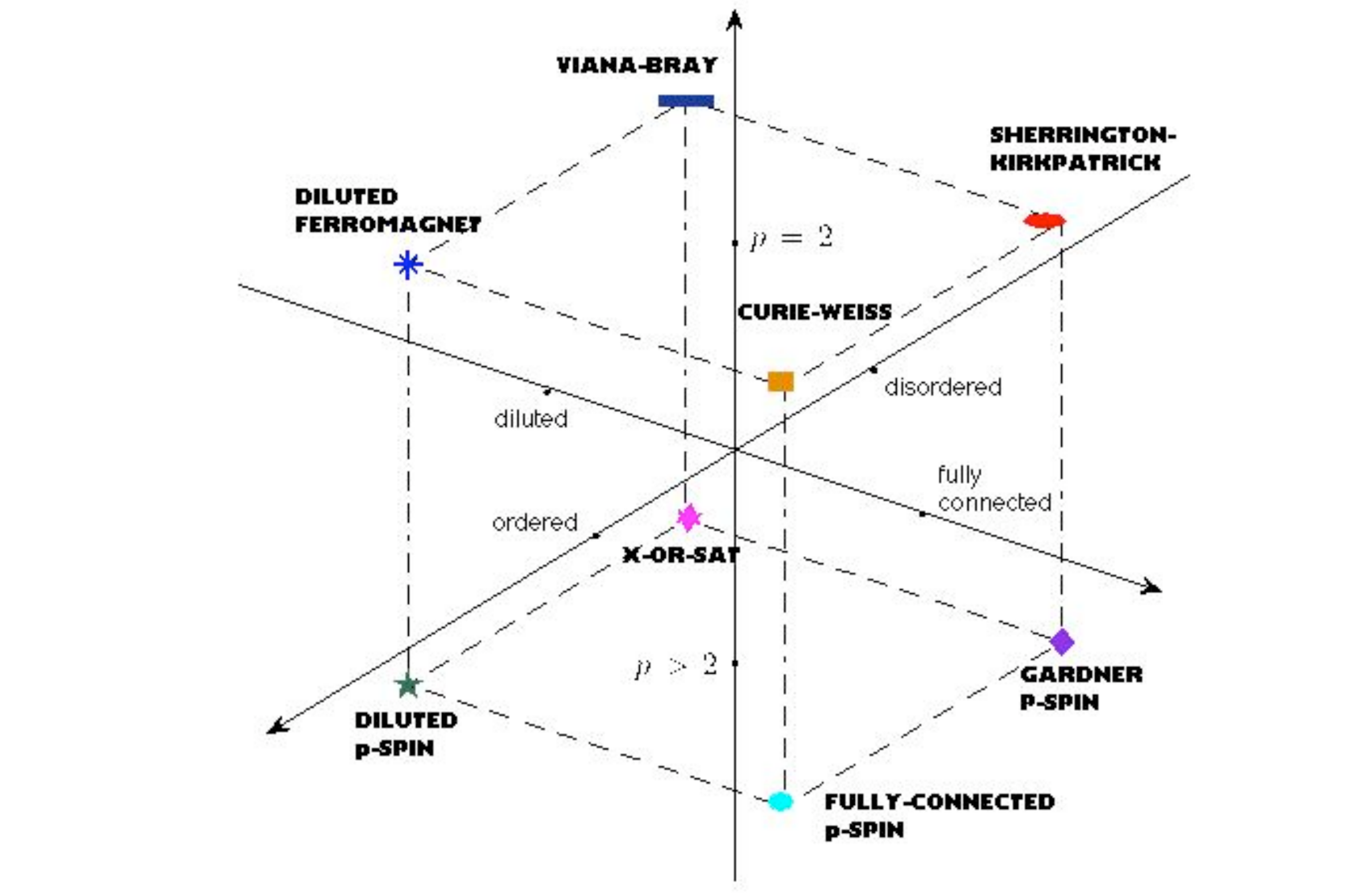}
\caption{\label{fig:cubo} Schematic representation of the connections among different models based on mean
field interactions between variables endowed with discrete symmetry.}
\end{figure}

In a nutshell, the system is a ferromagnet in which the
interactions happen in $p$-plets, instead of more classical couples,
and the interacting agents live on a diluted random network, i.e.
the Erd\"{o}s-Renyi graph. In general, the graph can be specified by fixing the
 number of nodes $N$ and its ``connectivity'' $\alpha$, which represents the number of nearest neighbors per site.

As standard ferromagnets, the model is shown to exhibit two
phases, a paramagnetic one and a (replica symmetric) ferromagnetic
one, on the the other hand, as a difference with respect to the
standard ferromagnet, the phase transition does not display
criticality for $p>2$. The model is investigated by means of
cavity field technique and extensive numerical simulations.

We find an expression for the free energy as a function of $p$, of
the network connectivity  $\alpha$ and of the (inverse)
temperature $\beta$, showing that it is consistent with known
results. In particular, by properly tuning $p$ and $\alpha$ we
recover the Curie-Weiss model \cite{abarra}, the diluted Ising
model \cite{ABC} and the fully-connected $p$-spin model
\cite{abarra3}; moreover, regardless of the (finite) dilution, for
$p=2$ criticality is restored. Full agreement with Monte Carlo
simulations is obtained both on the absence of the critical
behavior and on the free energy structure.

The paper is organized as follows: In Section $2$ the model is
introduced and some of its properties worked out together with the
introduction of a proper statistical mechanics machinery, while in
section $3$ its equilibrium is solved via the smooth cavity field
technique. Section $4$ deals with the properties of the free
energy and its consistency with well known models, while in
section $5$ our numerical analysis is presented. Section $6$ is
left for a summary and outlook. Finally, Section  $7$, as an
Appendix, contains the detailed proofs of the theorems introduced.

\section{The diluted even-$p$-spin ferromagnet}
In this section we explore the properties of a diluted even-$p$-spin ferromagnet: we restrict ourselves only to even values of
$p$ for mathematical convenience as the investigation with the
cavities is much simpler. However, due to monotonicity of all the
observables in $p$, such restriction does not imply any loss of
generality, as confirmed also by numerical simulations performed
on both even and odd values of $p$.

Before proceeding, it is worth recalling some concepts concerning
the diluted random network where the magnetic system is set. Such
a network is an Erd\"{o}s-Renyi graph \cite{ER} defined as
follows: given a number $N$ of nodes, we introduce connections
between them in such a way that each pair of vertices $i,j$ has a
connecting link with independent probability equal to $\alpha/N$,
with $0 \leq \alpha \leq N$. As a result, the probability
distribution for the number of links per node (or coordination
number) is binomial with average $\alpha$. Hence, the parameter
$\alpha$ provides a measure of the ``degree of connectivity'' of
the graph itself: the smaller $\alpha$ the more diluted the
system; for $\alpha = 0$ and $\alpha =N$ the extreme cases of
fully disconnected and fully connected graphs, respectively, are
recovered. Notice that in the thermodynamic limit $N \rightarrow
\infty$ the binomial distribution converges towards the Poisson
distribution \cite{BCC}.

The ER graph can be algebraically described by the so-called adjacency
matrix $\mathbf{A}$ which is an $N \times N$ symmetric matrix
whose entry $A_{ij}$ is $1$ if $i \neq j$ and the two nodes are
connected together, otherwise it is zero.


We now associate to each node $i$ a binary variable $\sigma_{i}=\pm1, \ i \in
[1,N]$ and we introduce $p$ families
$\{i_{\nu}^1\},\{i_{\nu}^2\},...,\{i_{\nu}^p\}$ of i.i.d. random
variables uniformly distributed on the previous interval. Then, the
Hamiltonian is given by the following expression
\begin{equation}\label{ham}
H_{N}(\sigma,\gamma)=-\sum_{\nu=1}^{k_{\gamma N}}
\sigma_{i_{\nu}^1}\sigma_{i_{\nu}^2}...\sigma_{i_{\nu}^p},
\end{equation}
where $k$ represents the number of connected $p$-plets present in
the graph. Reflecting the underlying network, $k$ is a Poisson
distributed random variable with mean value $\gamma N$. The
relation among the coordination number $\alpha$ and $\gamma$ is
$\gamma \propto \alpha^{p-1}$: this will be easily understood a
few lines later by a normalization argument coupled with the high
connectivity limit of this mean field model.

The quenched expectation of the model is given by the composition
of the Poissonian average with the uniform one performed over the
families $\{i_{\nu}\}$
\begin{equation}
\textbf{E}[\cdot] = E_PE_i[\cdot] = \sum_{k=0}^{\infty}
\frac{e^{-\gamma N}(\gamma
N)^k}{k!N^p}\sum_{i_{\nu}^1....i_{\nu}^p}^{1,N}[\cdot],
\end{equation}
where the term $N^p \approx N!/(N-p)!$ accounts for the number of
possible ordered $p$-plets.

As they will be useful in our derivation, it is worth stressing
the following properties of the Poisson distribution: Let us
consider a function $g:\mathbb{N}\to\mathbb{R}$, and a Poisson
variable $k$ with mean $\gamma N$, whose expectation is denoted by
$\mathbb{E}$.

It is easy to verify that
\begin{eqnarray}\label{Pp1}
\mathbb{E}[k g(k)] &=& \gamma N \mathbb{E} [g(k-1)] \\ \label{Pp2}
\partial_{\gamma N}\mathbb{E}[g(k)] &=&
\mathbb{E}[g(k+1)-g(k)]\\ \label{Pp3}
\partial^2_{(\gamma N)^2}\mathbb{E}[g(k)] &=&
\mathbb{E}[g(k+2)-2g(k+1)+g(k)].
\end{eqnarray}

The Hamiltonian written as in eq.~(\ref{ham}), has the advantage that it
is the sum of (a random number of) i.i.d.\ terms. To see the
connection to a more familiar Hamiltonian written in terms of
adjacency matrix elements, we first notice that being $\alpha /N$ the probability that two nodes are connected, among the $N^p$ possible
$p$-plets, the number of connected $p$-plets is
Poisson-distributed with average $\alpha^{p-1} N + O(\sqrt{N})$ for
large $N$. We now define the adjacency tensor $A_{i_1,...,i_p} \equiv A_{i_1,i_2}A_{i_1,i_3}...A_{i_1,i_p}$ which equals $1$ whenever the $p$-plet  $i_1,...,i_p$ occurs to be connected;  $A_{i_1,...,i_p}$ is Poisson distributed and has mean $\gamma N/N^{p} \sim (\alpha / N)^{p-1}$.
%
Hence, we can write the following Hamiltonian which is thermodynamically equivalent
to $H_N(\sigma,\gamma)$ appearing in eq.~(\ref{ham}):
\begin{equation}\label{parag}
-H_N(\sigma;  \gamma) \sim - \hat{H}_N(\sigma;\textbf{A}) = \sum_{i_1,...,i_p}^N
A_{i_1,...,i_p}\sigma_{i_1}...\sigma_{i_p}.
\end{equation}

Then, it is enough to consider the streaming of  the following
interpolating free energy (whose structure proves the statement a
priori by its thermodynamic meaning), depending on the real
parameter $t\in[0,1]$
$$
\phi(t) = \frac{\mathbb{E}}{N}\ln\sum_{\sigma}e^{\beta
\large(\sum_{\nu=1}^k \sigma_{i^1_{\nu}}...\sigma_{i^p_{\nu}} +
\sum_{i_1,...,i_p}^N A_{i_1,...,i_p}
\sigma_{i_1}...\sigma_{i_p}\large)},
$$
where $k$ is a Poisson random variable with mean $\gamma N t$ and
$A_{i_1,...,i_p}$ are random Poisson  variables with mean
$(1-t)\gamma/N^{p-1}$. In this way the two separated models are
recovered in the two extremals of the interpolation (for $t=0,1$).
By computing the $t$-derivative, we get
\begin{eqnarray}
\frac{1}{\gamma}\frac{d \phi(t)}{dt} &=& \mathbb{E}\ln(1+
\Omega(\sigma_{i_0^1}...\sigma_{i_0^p})\tanh(\beta))
\\ \nonumber &-& \frac{1}{N^{p}}\sum_{i_1,...,i_p}^N\ln(1+\Omega(\sigma_{i_1}...\sigma_{i_p})\tanh(\beta))=0,
\end{eqnarray}
where the label $0$ in $i_0^k$ stands for a new spin, born in the
derivative, according to the Poisson property (\ref{Pp2}); as
the $i_0$'s are independent of the random site indices in the
$t$-dependent $\Omega$ measure, the equivalence is proved.

\bigskip

Following a statistical mechanics approach, we know that the
macroscopic behavior, versus the connectivity $\alpha$ and the
inverse temperature  $\beta=1/T$, is described by the following
free energy density (often called {\itshape quenched pressure})
\begin{eqnarray}
A(\alpha,\beta) &=& \lim_{N \to \infty} A_N(\alpha,\beta) \\
\nonumber &=& \lim_{N \to \infty}\frac1N\textbf{E}\ln
Z_N(\gamma,\beta),\
\end{eqnarray}
where
\begin{equation}
Z_N(\gamma,\beta) = \sum_{\{\sigma\}}e^{-\beta H_N(\sigma,\gamma)}
\end{equation}
is the partition function. Taken $g(\sigma)$ as a generic function, the Boltzmann state is therefore given
by
\begin{equation}
\omega(g(\sigma)) = \frac{1}{Z_{N}(\gamma,\beta)}
\sum_{\{\sigma_N\}} g(\sigma) e^{-\beta H_{N}(\sigma,\gamma)},
\end{equation}
with its replicated form
\begin{equation}
\Omega(g(\sigma)) = \prod_s \omega^{(s)}(g(\sigma^{(s)}))
\end{equation}
and the total average $\langle g(\sigma)\rangle$ is defined as
\begin{equation}
\langle g(\sigma)\rangle = \textbf{E}[\Omega(g(\sigma))].
\end{equation}
\medskip
\newline
Let us introduce further, as order parameters of the theory, the
multi-overlaps \be
q_{1...n}=\frac1N\sum_{i=1}^{N}\sigma^{(1)}_{i}...\sigma^{(n)}_{i},
\ee with a particular attention at the magnetization $m = q_1
=(1/N)\sum_{i=1}^{N}\sigma_{i}$ and to the two replica overlap
$q_{12} = (1/N)\sum_{i=1}^{N}\sigma_{i}^1\sigma_i^2$.

The normalization constant of the quenched pressure can be checked
by performing the expectation value of the cost function:
\begin{eqnarray}\nonumber
\textbf{E}[H] &=& -\gamma N m^p \\  \textbf{E}[H^2] -
\textbf{E}^2[H] &=& \gamma^2 N^2\Big[(q_{12}^p - m^p ) + O \left(
\frac{1}{N} \right) \Big],
\end{eqnarray}
by which it is easy to see that the model is well defined, in
particular it is linearly extensive in the volume $N$. Then, in
the high connectivity limit each agent interacts with all the
others ($\alpha \sim N$) and, in the thermodynamic limit, $\alpha
\to \infty$. Now, such a high-connectivity limit, i.e. a linear
divergence of $\alpha$, is properly recovered for any finite $p$,
$p<N$. In particular, if $p=2$ the amount of couples in the
summation scales as $N(N-1)/2$ and $\gamma = 2\alpha$; if $p=3$
the amount of triples scales as $N(N-1)(N-2)/3!$ and, with $\gamma
= 3!\alpha^2$.

Before starting our free energy analysis, we want to point out
also  the connection between this diluted version and the fully
connected counterpart.
Let us remember that the Hamiltonian of the fully connected  $p$-spin
model (FC) can be written as \cite{abarra3} \be H^{FC}_{N}(\sigma)
= \frac{p!}{2 N^{p-1}}\sum_{1 \leq i_1 < ... < i_p \leq N}
\sigma_{i_1}\sigma_{i_2}...\sigma_{i_p}, \ee and let us consider
the  trial function $\hat{A}(t)$ defined as follows \be \hat{A}(t)
= \frac{1}{N}\mathbb{E}\ln \sum_{\sigma} \exp\Big[ \beta
\sum_{\nu}^{P_{\gamma N
t}}\sigma_{i_{\nu}^1}\sigma_{i_{\nu}^2}...\sigma_{i_{\nu}^p} +
(1-t)\frac{\beta' N}{2}m^p \Big], \ee which interpolates between
the fully connected  $p$-spin model and the diluted one, such that
for $t=0$ only the fully connected survives, while the opposite
happens for $t=1$. Let us work out the derivative with respect to
$t$ to obtain \begin{eqnarray} \partial_t \hat{A}(t) &=&
(p-1)\alpha^{p-1}\ln\cosh(\beta) \\ \nonumber &-& (p-1)
\alpha^{p-1}\sum_{n}\frac{-1^n}{n}\theta^n \langle q_n^p\rangle -
\frac{\beta'}{2}\langle m^p \rangle, \end{eqnarray} by which we
see that the correct scaling, in order to recover the proper
infinite connectivity model, is obtained when $\alpha \to \infty$,
$\beta \to 0$ and $\beta' = 2(p-1) \alpha^{p-1}\tanh(\beta)$ is
held constant.

\begin{remark}
It is worth noting that for $p=2$ we recover the correct scaling
of the diluted Curie-Weiss model \cite{ABC}, furthermore the
dilute $p$-spin model reduces to the fully connected one, in the
infinite connectivity limit, uniformly in the size of the system.
\end{remark}

\section{The smooth cavity approach}

In this section we want to look for an iterative expression of the
free energy density by using a version of the cavity strategy
\cite{barra1,abarra} that we briefly recall: the idea behind the
cavity techniques \cite{guerra2,MPV}, which, for our purposes,
resembles the stochastic stability approach \cite{cg2,parisiSS},
is that information concerning the free energy density can be
extrapolated when looking at the incremental extensive free energy
given by the addition of a spin.
\newline
In diluted models, this additional spin changes also (infinitesimally
in the high $N$ limit) the connectivity and, in evaluating how the
free energy density varies conformingly with this, we are going to
prove that it can be written by a cavity function and such a
connectivity shift.
\newline
So the behavior of the system is encoded into these two parts. The
latter is simpler as it is made up only by stochastically stable
terms (a proper definition of these terms will follow in the current
section). Conversely, the former term needs to be expressed via these
terms and this must be achieved by iterative expansions.

At first we show how the free energy density can be decomposed via
these two parts (the cavity function and the connectivity shift).
Then, we analyze each term separately. We will see that they can
be expressed by the momenta of the magnetization and of the
multi-overlaps, weighted in a perturbed Boltzmann state, which
recovers the standard one in the thermodynamic limit.
\begin{theorem}\label{primolevi}
\textit{In the thermodynamic limit, the quenched pressure of the
even  $p$-spin diluted ferromagnetic model is given by the
following expression} \be A(\alpha,\beta) = \ln2
-\frac{\alpha}{p-1}\frac{d}{d\alpha}A(\alpha,\beta) +
\Psi(\alpha,\beta,t=1), \ee
\end{theorem}
where the cavity function $\Psi(t,\alpha,\beta)$ is introduced as
\begin{eqnarray}\label{cavity1}
&& \textbf{E}\Big[\ln\frac{\sum_{\{\sigma\}}
e^{\beta\sum_{\nu=1}^{k_{\tilde{\gamma}N}}
\sigma_{i_{\nu}^1}\sigma_{i_{\nu}^2}...\sigma_{i_{\nu}^p}}\;
e^{\beta\sum_{\nu=1}^{k_{2\tilde{\gamma}t}}
\sigma_{i_{\nu}^1}\sigma_{i_{\nu}^2}...\sigma_{i_{\nu}^{p-1}}}}
{\sum_{\{\sigma\}} e^{\beta\sum_{\nu=1}^{k_{\tilde{\gamma}N}}
\sigma_{i_{\nu}^1}\sigma_{i_{\nu}^2}...\sigma_{i_{\nu}^p}}}\Big]= \nonumber \\
&&  \textbf{E}\Big[\ln \frac{Z_{N,t}(\tilde{\gamma},\beta)}
{Z_{N}(\tilde{\gamma},\beta)}\Big] =
\Psi_N(\tilde{\gamma},\beta,t),
\end{eqnarray}
with \be\label{cavity2} \Psi(\gamma,\beta,t) =
\lim_{N\rightarrow\infty}\Psi_N(\tilde{\gamma},\beta,t). \ee
\medskip
\newline
For the sake of clearness and to avoid interrupting the paper with
long technical calculations, the proof of the theorem is reported
in the Appendix.
\newline
Thanks to the previous theorem, it is possible to figure out an
expression for the pressure by studying the properties of the
cavity function $\Psi(\alpha,\beta)$ and the connectivity shift
$\partial_{\alpha}A(\alpha,\beta)$.
\newline
Using the properties of the Poisson distribution (\ref{Pp1},
\ref{Pp2}), we can write
\begin{eqnarray}
\frac{d}{d\alpha}A(\alpha,\beta)
&=& \frac{(p-1)}{N}\alpha^{p-2}\frac{d}{d\gamma}\textbf{E}\Big[\ln Z_N(\gamma,\beta)\Big] = \nonumber \\
&=& (p-1)\alpha^{p-2}\textbf{E}\Big[\ln \sum_{\{\sigma\}}
e^{\beta\sum_{\nu=1}^{k+1}
\sigma_{i_{\nu}^1}...\sigma_{i_{\nu}^p}} -  \nonumber \\ \nonumber
&-& \ln \sum_{\{\sigma\}} e^{\beta\sum_{\nu=1}^{k}
\sigma_{i_{\nu}^1}...\sigma_{i_{\nu}^p}}\Big]. \nonumber
\end{eqnarray}
Now considering the relation (and definition)
\begin{eqnarray}
e^{\beta\sigma_{i_0^1}...\sigma_{i_0^p}} &=&
\cosh\beta + \sigma_{i_0^1}...\sigma_{i_0^p}\sinh\beta, \\
\theta &=& \tanh\beta,
\end{eqnarray}
we can write
\begin{eqnarray}
&& \frac{d}{d\alpha}A(\alpha,\beta) = \\ \nonumber &&
(p-1)\alpha^{p-2}\Big[\ln\cosh\beta + \textbf{E}[\ln(1 +
\omega(\sigma_{i_{\nu}^1}...\sigma_{i_{\nu}^p})\theta)]\Big].
\end{eqnarray}
At the end, expanding the logarithm, we obtain
\begin{eqnarray}
\frac{d}{d\alpha}A(\alpha,\beta) &=&
(p-1)\alpha^{p-2}\ln\cosh\beta - \\ \nonumber
 &-& (p-1)\alpha^{p-2}\sum_{n=1}^{\infty}\frac{(-1)^n}{n}\theta^n
\langle q_{1,...,n}^p \rangle.
\end{eqnarray}
\medskip
\newline
With the same procedure it is possible to show that
\begin{eqnarray}\label{Psit}
\frac{d}{dt}\Psi(\tilde{\alpha},\beta,t) &=&
2\tilde{\alpha}^{p-1}\ln\cosh\beta - \\ \nonumber &-&
2\tilde{\alpha}^{p-1}\sum_{n=1}^{\infty}\frac{(-1)^n}{n}\theta^n
\langle q_{1,...,n}^{p-1} \rangle_{\tilde{\alpha},t},
\end{eqnarray}
where
$$
\tilde{\alpha} = \alpha \Big[\frac{N}{N+1}\Big]^{\frac{1}{p-1}}
\qquad.
$$
Now, by eq.~(\ref{Psit}), we see that
even the cavity function, once the r.h.s. of eq.(\ref{Psit}) is
integrated back against $t$, can be expressed via all the order
parameters of the model:
\begin{equation}\nonumber
\Psi(\tilde{\alpha},\beta,t)=2\tilde{\alpha}^{p-1}\Big(\ln\cosh(\beta)-
\sum_{n=1}^{\infty}\frac{(-\theta)^n}{n}\int_0^t \langle
q_{1,...,n}^{p-1}\rangle_{\tilde{\alpha},t} \Big).
\end{equation}

So, as expected, we can understand the properties of the free
energy by analyzing the properties of the order parameters:
magnetization and overlaps, weighted in their extended Boltzmann
state $\tilde{\omega}_t$.
\newline
Further, as we expect that the order parameters is able to describe
thermodynamics even in the true Boltzmann states $\omega, \Omega$
\cite{landau}, accordingly to the following definitions, we are
going to show that {\em filled} order parameters (the ones
involving even numbers of replicas) are stochastically stable or,
in other words, are independent of the $t$-perturbation in the
thermodynamic limit, while the others, not filled, become filled,
again in this limit (such that even for them $\omega_t\to \omega$
in the high $N$ limit and thermodynamics is recovered). The whole
is explained in the following definitions and theorems of this
section.

\begin{definition}
We define the t-dependent Boltzmann state $\tilde{\omega}_t$ as
\begin{eqnarray}\label{dente}
&& \tilde{\omega}_t(g(\sigma)) = \\ \nonumber &&
\frac{1}{Z_{N,t}(\gamma,\beta)} \sum_{\{\sigma\}}g(\sigma)
e^{\beta\sum_{\nu=1}^{k_{\tilde{\gamma}N}}
\sigma_{i_{\nu}^1}...\sigma_{i_{\nu}^p} +
\beta\sum_{\nu=1}^{k_{2\tilde{\gamma}t}}
\sigma_{i_{\nu}^1}...\sigma_{i_{\nu}^{p-1}}}\label{dante},
\end{eqnarray}
where $Z_{N,t}(\gamma,\beta)$ extends the classical partition
function in the same spirit of the numerator of eq.(\ref{dente})
itself, and $\tilde{\gamma} = \gamma(1+N^{-1})$.
\end{definition}
We see that the original Boltzmann state of a $N$-spin system is
recovered as $t$ approached $0$, while, in the limit $t \to 1$ and
gauging the spins, it is possible to build a Boltzmann state of a
$N+1$ spins, with a little shift both in $\alpha,\beta$, which
vanishes in the $N \to \infty$ limit.

Now, coherently with the implication of thermodynamic limit (by which
$A_{N+1}(\alpha,\beta)-A_N(\alpha,\beta) =0$ for $N \to \infty$),
we are going to define the {\em filled} overlap monomials and show
their independence (stochastic stability) with respect to the
perturbation encoded by the interpolating parameter $t$.
These parameters are already `good" order parameters describing
the theory, while the others (the not-filled ones) must be
expressed via the formers, and this will be achieved by expanding
them with a suitably introduced streaming equation.

\begin{definition} We can split the class of monomials of the
order parameters in two families:
\begin{itemize}

\item We define {\itshape filled} or equivalently {\itshape stochastically
stable} those overlap monomials with all the replicas appearing an
even number of times (i.e. $q_{12}^2$,\ $m^2$,\
$q_{12}q_{34}q_{1234}$).

\item We define {\itshape non-filled}
those overlap monomials with at least one replica appearing an odd
number of times (i.e. $q_{12}$,\ $m$,\ $q_{12}q_{34}$).
\end{itemize}
\end{definition}
We are going to show three theorems that will play a guiding role
for our iteration: as this approach has been deeply developed in
similar contexts (as fully connected Ising and  $p$-spin models
\cite{abarra,abarra3}, fully connected spin glasses \cite{barra1}
or diluted ferromagnetic models \cite{ABC,BCC}, which are the
``boundaries'' of the model of this paper) we will not show all
the details of the proof, but we sketch them in the appendix as
they are really intuitive. The interested reader will found a
clear derivation in the appendix and can deepen this point by
looking at the original works.
\begin{theorem}\label{ciccia}
In the thermodynamic limit and setting $t=1$ we have \be
\tilde{\omega}_{N,t}(\sigma_{i_1}\sigma_{i_2}...\sigma_{i_n}) =
\tilde{\omega}_{N+1}(\sigma_{i_1}\sigma_{i_2}...\sigma_{i_n}\sigma_{N+1}^n).
\ee
\end{theorem}
\begin{theorem}\label{saturabili}
Let $Q_{ab}$ be a not-filled monomial of the overlaps (this means
that $q_{ab}Q_{ab}$ is filled). We have \be
\lim_{N\rightarrow\infty}\lim_{t\rightarrow1} \langle Q_{ab}
\rangle_t = \langle q_{ab}Q_{ab} \rangle, \ee (examples:
\newline for $N \rightarrow \infty$ we get $\langle m_1 \rangle_t
\rightarrow \langle m_1^2 \rangle,\quad \langle q_{12} \rangle_t
\rightarrow \langle q_{12}^2 \rangle$).
\end{theorem}
\begin{theorem}\label{saturi}
In the $N\rightarrow\infty$ limit, the averages
$\langle\cdot\rangle$ of the filled polynomials are t-independent
in $\beta$ average.
\end{theorem}
\medskip

\section{Properties of the free energy}
In this section we are going to address various points: at first
we work out the constraints that the model must fulfil, which are
in agreement both with a self-averaging behavior of the
magnetization and with the replica-symmetric behavior of the
multi-overlaps \cite{gulielmo}; then we write an iterative
expression for the free energy density and its links with known
models as diluted ferromagnets ($p\to2$ limit) and fully connected
$p$-spin models ($\alpha \to \infty$ limit).

With the following definition \begin{eqnarray} \tilde{\beta} &=&
2(p-1)\tilde{\alpha}^{p-1}\theta \\ \nonumber &=&
2(p-1)\alpha^{p-1}\frac{N}{N+1}\theta \quad
\stackrel{N\rightarrow\infty}{\longrightarrow}
2(p-1)\alpha^{p-1}\theta = \beta', \end{eqnarray} we show (and
prove in the Appendix) the streaming of replica functions, by
which not filled multi-overlaps can be expressed via filled ones.
\begin{proposition}\label{stream}
Let $F_s$ be a function of s replicas. Then the following
streaming equation holds
\begin{eqnarray}
\frac{\partial\langle F_s \rangle_{t,\tilde{\alpha}}}{\partial t}
&=& \tilde{\beta} \Big[\sum_{a=1}^s\langle F_s
m_a^{p-1}\rangle_{t,\tilde{\alpha}} - s \langle F_s
m_{s+1}^{p-1}\rangle_{t,\tilde{\alpha}}\Big] \quad
\\ \nonumber
&+& \tilde{\beta}\theta \Big[ \sum_{a<b}^{1,s}\langle F_s
q_{a,b}^{p-1} \rangle_{t,\tilde{\alpha}} - s\sum_{a=1}^s\langle
F_s q_{a,s+1}^{p-1}\rangle_{t,\tilde{\alpha}} \\ \nonumber &+&
\frac{s(s+1)}{2!}\langle F_s
q_{s+1,s+2}^{p-1}\rangle_{t,\tilde{\alpha}}\Big] + O(\theta^2).
\end{eqnarray}
\end{proposition}
\begin{remark}
We stress that, at the first two level of approximation presented
here, the streaming has the structure of a $\theta$-weighted
linear sum of the Curie-Weiss streaming ($\theta^0$ term)
\cite{abarra} and the Sherrington-Kirkpatrick streaming
($\theta^1$ term) \cite{barra1}, providing mathematical structures of disordered systems with a certain degree of
independence with respect to the kind of quenched noise (frustration or
dilution).
\end{remark}

\bigskip

It is now immediate to obtain the linear order parameter
constraints (often known as Aizenman-Contucci polynomials
\cite{ac,abarra,BCC}) of the theory: in fact, the generator of
such a constraint is the streaming equation when applied on each
filled overlap monomial (or equivalently it is possible to apply
the streaming on a not-filled one and then gauge the obtained
expression; for the sake of clearness both the methods will be
exploited, the former for $q_2$ and the latter for $m$).

As examples, dealing with the terms $m^{p-1}$ and $q_{2}^{p-1}$,
it is straightforward to check that \begin{eqnarray}\nonumber 0
&=& \lim_{N \to \infty} \frac{\partial\langle m_N^{p-1}
\rangle_{t,\tilde{\alpha}}}{\partial t} = \tilde{\beta} \Big(
\langle m_1^{2(p-1)} \rangle - \langle m_1^{p-1} \rangle^2 \Big)\\
\nonumber &+& \tilde{\beta}\theta \Big( \langle m_1^{p-1}q_2^{p-1}
\rangle - \langle m_1^{p-1}\rangle \langle  q_2^{p-1} \rangle
\Big) + O(\theta^3), \nonumber \end{eqnarray} then, by gauging the
above expression, in the thermodynamic limit, (as $\lim_{N \to
\infty}\langle m_N^{p-1}\rangle_t \to \langle m^p \rangle$), we
get
$$
\Big( (\langle m_1^{2p} \rangle - \langle m_1^p \rangle^2) +
\theta ( \langle q_{2}^{2p} \rangle - \langle q_{2}^p \rangle^2
)\Big)=0, \ \ \forall \theta \in \mathbb{\mathcal{R}}^+.
$$
The fact that the previous expression holds for every $\theta$
suggests self-averaging for the energy (by which all the linear
constraints can be derived \cite{BCC}) due to the first term, as
well as replica symmetric behavior of the two replica overlap due
to the last one. Analogously, the contribution of the $\langle q_{2}^2\rangle$
generator is  \begin{eqnarray}  \nonumber 0 &=& \Big( (\langle
q_{12}^{p-1} m_1^{p-1} \rangle + \langle q_{12}^{p-1} m_2^{p-1}
\rangle - 2 \langle q_{12}^{p-1} m_3^{p-1} \rangle) +  \\
\nonumber &+& \theta ( \langle q_{12}^{p-1} q_{12}^{p-1} \rangle -
4 \langle q_{12}^{p-1} q_{23}^{p-1} \rangle + 3 \langle
q_{12}^{p-1} q_{34}^{p-1} \rangle) \Big), \end{eqnarray} which
shows replica symmetric behavior of the magnetization by the first
term and the classical Aizenman-Contucci relations \cite{ac,BCC}
by the latter.

Furthermore, turning now our attention to the free energy, it is
easy to see that the streaming equation  allows to generate all
the desired overlap functions coupled to every well behaved $F_s$.
In this way, if $F_s$ is a not filled overlap, we can always
expand recursively it into a filled one, the only price to pay
given by the $\theta$ order that has to be reached or, which is
equivalent, the number of derivatives that have to be performed.

Let us now remember the t-derivative of the cavity function
(\ref{Psit}), showing explicitly the first two terms of its
expansion
\begin{eqnarray}\label{Psit2}
\frac{d}{dt}\Psi(\tilde{\alpha},\beta,t) &=&
2\tilde{\alpha}^{p-1}\ln\cosh\beta + \tilde{\beta}\langle
m_1^{p-1} \rangle_{\tilde{\alpha},t} - \\ \nonumber  &-&
\frac{\tilde{\beta}}{2}\theta\langle q_{12}^{p-1}
\rangle_{\tilde{\alpha},t} -
2\tilde{\beta}^{p-1}\sum_{n=3}^{\infty}\frac{-1^n\theta^n}{n}
\langle q_{1,...,n}^{p-1} \rangle_{\tilde{\alpha},t}.
\end{eqnarray}

As derivative of fillable terms involve filled ones, we can arrive
to an analytical form of $\Psi(\alpha,\beta)$ if we calculate it
as the $t$-integral of its $t$-derivative, together with the
obvious relation $\Psi(t=0) = 0$. Hence, if we apply the streaming
equation machinery to the overlaps constituting equation
(\ref{Psit2}), we are able to fill them and to remove their
$t$-dependence in the thermodynamic limit. In this way we are
allowed to bring them out from the final $t$-integral.
\newline
In fact, without gauging (so, not only in the ergodic regime,
where symmetries are preserved), we can expand the streaming of
$\langle m^{p-1}\rangle_t$:
\newline
\begin{eqnarray}
\frac{d\langle m_1^{p-1} \rangle_t}{dt} &=&
\tilde{\beta}\Big[\langle m_1^{2(p-1)} \rangle - \langle m_1^{p-1}m_2^{p-1} \rangle_t\Big] + \nonumber \\
&-&  \tilde{\beta}\theta\Big[\langle m_1^{p-1}q_{12}^{p-1}
\rangle_t - \langle m_1^{p-1}q_{23}^{p-1} \rangle_t\Big] +
O(\theta^2). \nonumber
\end{eqnarray}
We can note the presence of the filled monomial $\langle
m_1^{2(p-1)} \rangle$, whose $t$-dependence has been omitted
explicitly to underline its stochastic stability, while the overlaps
$\langle m_1^{p-1}m_2^{p-1} \rangle_t$ and $\langle
m_1^{p-1}q_{12}^{p-1} \rangle_t$ can be saturated in two steps of
streaming. This will be sufficient, wishing to have a fourth order
expansion for the cavity function.
\newline
We now derive these two functions and apply the same scheme to all
the overlaps that appear and that have to be necessary filled in
order to obtain the desired result.
\begin{eqnarray}
&& \frac{d\langle m_1^{p-1}m_2^{p-1} \rangle_t}{dt} = \nonumber \\
&& 2\tilde{\beta}\Big[
\langle m_1^{2(p-1)}m_2^{p-1} \rangle_t - \langle m_1^{p-1}m_2^{p-1}m_3^{p-1} \rangle_t\Big] + \nonumber \\
&& \theta \tilde{\beta} \Big[\langle
m_1^{p-1}m_2^{p-1}q_{12}^{p-1} \rangle -
4\langle m_1^{p-1}m_2^{p-1}q_{13}^{p-1} \rangle_t +  \nonumber \\
&&  3\langle m_1^{p-1}m_2^{p-1}q_{34}^{p-1} \rangle_t\Big],
\label{m1m2}
\end{eqnarray}
\begin{eqnarray}
&& \frac{d\langle m_1^{2(p-1)}m_2^{p-1} \rangle_t}{dt} = \\
\nonumber && 2\tilde{\beta}\Big[ \langle m_1^{2(p-1)}m_2^{2(p-1)}
\rangle_t\Big] +\tilde{\beta}\Big[\mbox{unfilled terms}\Big] +
O(\theta^2).
\end{eqnarray}
Integrating back in $t$ and neglecting higher order terms we have
\begin{equation}
\langle m_1^{2(p-1)}m_2^{p-1} \rangle_t =
\tilde{\beta}\Big[\langle m_1^{2(p-1)}m_2^{2(p-1)} \rangle\Big]t,
\end{equation}
and we can write
\begin{eqnarray}
&& \langle m_1^{p-1}m_2^{p-1} \rangle_t = \\ \nonumber &&
\tilde{\beta}\theta\langle m_1^{p-1}m_2^{p-1}q_{12}^{p-1} \rangle
t + \tilde{\beta}^2\langle m_1^{2(p-1)}m_2^{2(p-1)} \rangle t^2.
\end{eqnarray}
Let us take a look now at the other overlap $\langle
m_1^{p-1}q_{12}^{p-1} \rangle_t$:
\begin{eqnarray}\label{m1m2}
\frac{d\langle m_1^{p-1}q_{12}^{p-1} \rangle_t}{dt} &=&
\tilde{\beta}\Big[
\langle m_1^{2(p-1)}q_{12}^{p-1} \rangle_t - \langle m_1^{p-1}m_2^{p-1}q_{12}^{p-1} \rangle_t  \nonumber \\
&-& 2\langle m_1^{p-1}m_2^{p-1}m_3^{p-1}q_{12}^{p-1}
\rangle_t\Big] + O(\theta^2),
\end{eqnarray}
that gives
\begin{equation}
\langle m_1^{p-1}q_{12}^{p-1} \rangle_t = \tilde{\beta}\langle
m_1^{p-1}m_2^{p-1}q_{12}^{p-1} \rangle t + O(\theta^2).
\end{equation}

At this point we can write for $\langle
m_1^{p-1}\rangle_{t,\tilde{\alpha}}$ (and consequently for
$\langle q_{12}^{p-1}\rangle_{t,\tilde{\alpha}}$)
\begin{eqnarray}\nonumber
\langle m_1^{p-1}\rangle_{t,\tilde{\alpha}} &=&
\tilde{\beta}\langle m_1^{2(p-1)}\rangle t -
\frac{\tilde{\beta}^3}{3}\langle m_1^{2(p-1)}m_2^{2(p-1)}\rangle
t^3 \\ \nonumber &-&
\tilde{\beta}^2\theta\langle m_1^{p-1}m_2^{p-1}q_{12}^{p-1}\rangle t^2 + O(\theta^3), \nonumber \\
\langle q_{12}^{p-1}\rangle_{t,\tilde{\alpha}} &=&
\tilde{\beta}\theta\langle q_{12}^{2(p-1)}\rangle t +
\tilde{\beta}^2\langle m_1^{p-1}m_2^{p-1}q_{12}^{p-1}\rangle t^2 +
O(\theta^3).\nonumber
\end{eqnarray}
With these relations, eq. (\ref{Psit2}) becomes
\begin{eqnarray}\nonumber
&& \frac{d}{dt}\Psi_N(\alpha,\beta,t) = 2\alpha^{p-1}\ln\cosh\beta
+ \tilde{\beta}^2\langle m_1^{2(p-1)}\rangle t  \\ \nonumber && \
\ \ \  -\frac{\tilde{\beta}^2\theta^2}{2}\langle
q_{12}^{2(p-1)}\rangle t - \frac{3\tilde{\beta}^3\theta}{2}
\langle m_1^{p-1}m_2^{p-1}q_{12}^{p-1}\rangle t^2 \\ \nonumber &&
\ \ \ \  - \frac{\tilde{\beta}^4}{3}\langle
m_1^{2(p-1)}m_2^{2(p-1)}\rangle t^3 + O(\theta^5),\nonumber
\end{eqnarray}
which  ultimately allows us to write an iterated expressions for
$\Psi$ evaluated at $t=1$
\begin{eqnarray}
&& \Psi_N(\alpha,\beta,1) = \\ \nonumber &&
2\alpha^{p-1}\ln\cosh\beta + \frac{\tilde{\beta}^2}{2}\langle
m_1^{2(p-1)}\rangle  -
\frac{\tilde{\beta}^2\theta^2}{4}\langle q_{12}^{2(p-1)}\rangle - \\
&& \frac{\tilde{\beta}^3\theta}{2} \langle
m_1^{p-1}m_2^{p-1}q_{12}^{p-1}\rangle -
\frac{\tilde{\beta}^4}{12}\langle m_1^{2(p-1)}m_2^{2(p-1)}\rangle
t^3 + O(\theta^5). \nonumber
\end{eqnarray}
Overall the result we were looking for, namely a Landau-like polynomial
form for the free energy, reads off as
\begin{eqnarray}\label{maino}
A(\alpha,\beta) &=& \ln2 \:+\: \alpha^{p-1}\ln\cosh\beta +\\
&+& \frac{\beta'}{2}\Big(\beta'\langle m^{2(p-1)}\rangle - \langle
m^{p}\rangle\Big) + \nonumber \\ \nonumber &+&
\frac{\beta'\theta}{4}\Big(\beta'\theta\langle
q_{12}^{2(p-1)}\rangle - \langle q_{12}^{p}\rangle\Big) +
O(\theta^5). \nonumber
\end{eqnarray}

Now, several conclusions can be addressed from the expression
(\ref{maino}):
\newline
In fact, as we are going to see immediately through remarks, this
formula can bridge free-energies of quite different models
(diluted versus non-diluted, critical versus uncritical) and acts
as a general free energy expression close to the phase transition.
\begin{remark}
At first let us note that, by constraining the interaction to be
pairwise, critical behavior should arise \cite{landau}.
Coherently, we see that for $p=2$ we can write the free energy
expansion as
$$A(\alpha,\beta)_{p=2} = \ln 2 + \alpha
\ln\cosh(\beta) - \frac{\beta'}{2}(1-\beta')\langle m^2 \rangle -
\frac{\beta' \theta}{4} \langle q_2^2 \rangle,$$ which coincides
with the one of the diluted ferromagnet \cite{ABC} and displays
criticality at $2\alpha\theta=1$, where the coefficient of the
second order term vanishes, in agreement with previous results
\cite{ABC} and Landau theory \cite{landau}.
\end{remark}
\begin{remark}
The free energy density of the fully connected $p$-spin model
is \cite{abarra3} $A(\beta')= \ln 2 + \ln\cosh(\beta m^{p-1}) -
(\beta/2)m^p$, which coincides with the expansion (\ref{maino}) in
the limit of $\alpha \to \infty$ and $\beta \to 0$ with $\beta' =
2(p-1) \alpha^{p-1}\theta$.
\end{remark}
\begin{remark}
It is worth noting that the connectivity no longer plays a linear
role in contributing to the free energy density, as it does happen
for the diluted two body models \cite{ABC,gt2}. This is
interesting in applications to economic networks, where, for high
values of coordination number it may be interesting to develop
strategies with more than one coupling \cite{cicciona}.
\end{remark}

\section{Numerics}\label{sec:numerics}
We now analyze the system described in the previous section, from
the numerical point of view by performing extensive Monte Carlo
simulations. Within this approach it is more convenient to use the
second Hamiltonian introduced (see eq.(\ref{parag})):
\begin{equation}\label{my_hamiltonian}
\hat{H}_N(\sigma, \mathbf{A})=-\sum_{i_i}^{N}
\sigma_{i_1}\sum_{i_2<i_3<...<i_p=1}^{N} A_{i_1,...,i_p}
\sigma_{i_{2}}\sigma_{i_{3}}...\sigma_{i_{p}}.
\end{equation}
The product between the elements of the adjacency tensor ensures
that the $p-1$ spins considered in the second sum are joined by a
link with $i_1$.
\newline
The evolution of the magnetic system is realized by means of a
single spin-flip dynamics based on the Metropolis algorithm
\cite{barkema}. At each time step a spin is randomly extracted and
updated whenever its coordination number is larger than $p-1$. For
$\alpha$ large enough (at least above the percolation threshold,
as  obviously holds for the results found previously) and $p=3,4$
this condition is generally verified. The updating procedure for a
spin $\sigma_i$ works as follows: Firstly we calculate the energy
variation $\Delta E_i$ due to a possible spin flip, which for
$p=3$ and $p=4$ reads respectively
\begin{eqnarray}
\Delta E_i &=& 2 \sigma_i \sum_{j<k=1}^{N} A_{i,j}A_{i,k}
\sigma_{j}\sigma_{k},
\\
\Delta E_i &=& 2 \sigma_i \sum_{j<k<w=1}^{N} A_{i,j}A_{i,k}A_{i,w}
\sigma_{j}\sigma_{k}\sigma_{w}.
\end{eqnarray}
Now, if $\Delta E_i <0$, the spin-flip $\sigma_i \rightarrow -
\sigma_i$ is realized with probability $1$, otherwise it is
realized with probability $e^{-\beta \Delta E}$.

The cases $p=3, 4$ were studied in details, while  for $p=2$ we refer to
\cite{ABC}. Our investigations are aimed to evidence the existence of a phase transition and its nature and also to highlight a proper scaling for the temperature as the
parameter $\alpha$ is tuned.

As for the first point, we measured the so-called Binder cumulants
defined as follows:
\begin{equation}
G_N(T(\alpha)) \equiv 1 - \frac{ \langle m^4 \rangle_N }{3  \langle m^2 \rangle_N ^2},
\end{equation}
where $\langle \cdot \rangle_N$ indicates the statistical average obtained for a system of size $N$ and $T=\beta^{-1}$ \cite{binder}. The study of Binder
cumulants is particularly useful to locate and catalogue the
phase transition. In fact, in the case of continuous phase
transitions, $G_N(T)$ takes a universal positive value at the
critical point $T_c$, namely all the curves obtained for different
system sizes $N$ cross each other. On the other hand, for a
first-order transition $G_N(T)$ exhibits a minimum at $T_{min}$,
whose magnitude diverges as $N$. Moreover, a crossing point at
$T_{cross}$ can be as well detected when curves pertaining to
different sizes $N$ are considered \cite{vollmayr}. Now, $T_{min}$
and $T_{cross}$ scale as $T_{min}-T_c \propto N^{-1}$ and
$T_{cross}-T_c \propto N^{-2}$, respectively.

\begin{figure}[tb]
\bc
\includegraphics[height=60mm]{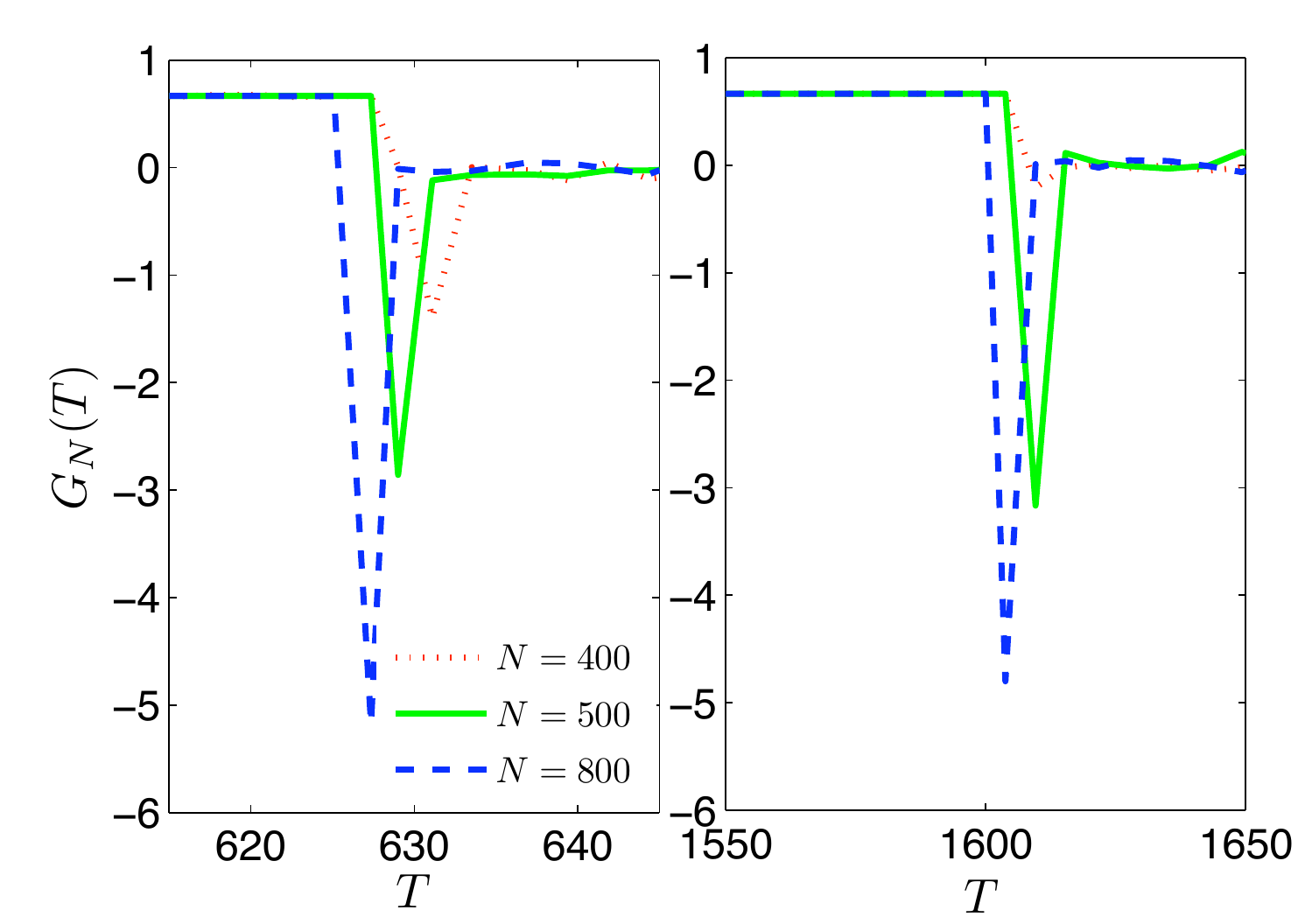}
\caption{\label{fig:Binder} Binder cumulants $G_L(T)$ for systems with $p=3$ and different size
$N$, as shown in the legend, and connectivity
$\alpha=50$ (left panel) and $alpha=80$ (right panel).}
\ec
\end{figure}

In Fig.~\ref{fig:Binder} we show data for $G_N(T)$ obtained for
systems of different sizes ($N=400$, $N=500$, and $N=800$) but
equal connectivity ($\alpha=50$ and $\alpha=80$, respectively) as
a function of the temperature $T$. The existence of a minimum is
clear and it occurs at $T \approx 625$ and $T \approx 1600$.
Similar results are found also for $p=4$ and they all highlight
the existence of a first-order phase transition (hence lack of
criticality) at a temperature which depends on the connectivity
$\alpha$.

\begin{figure}[tb]\bc
\includegraphics[height=60mm]{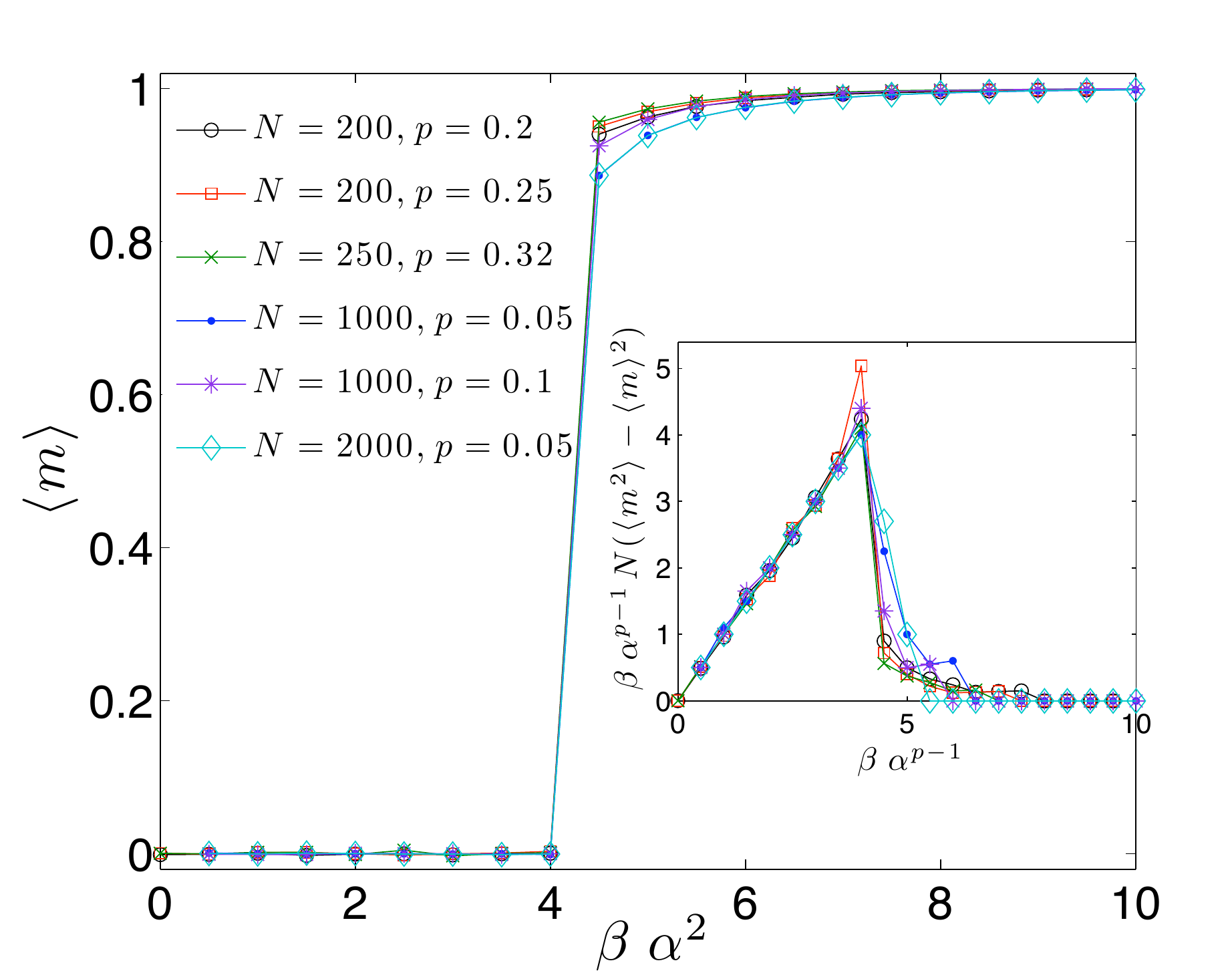}
\caption{\label{fig:p3}Magnetization (main figure) and its
normalized fluctuations (inset) for $3$-spin systems of different sizes and
different dilution as a function of $\beta \; \alpha^{p-1}$. The
collapse of all the curves provides a strong evidence for the
scaling of the temperature.} \ec
\end{figure}

\begin{figure}[tb]\bc
\includegraphics[height=60mm]{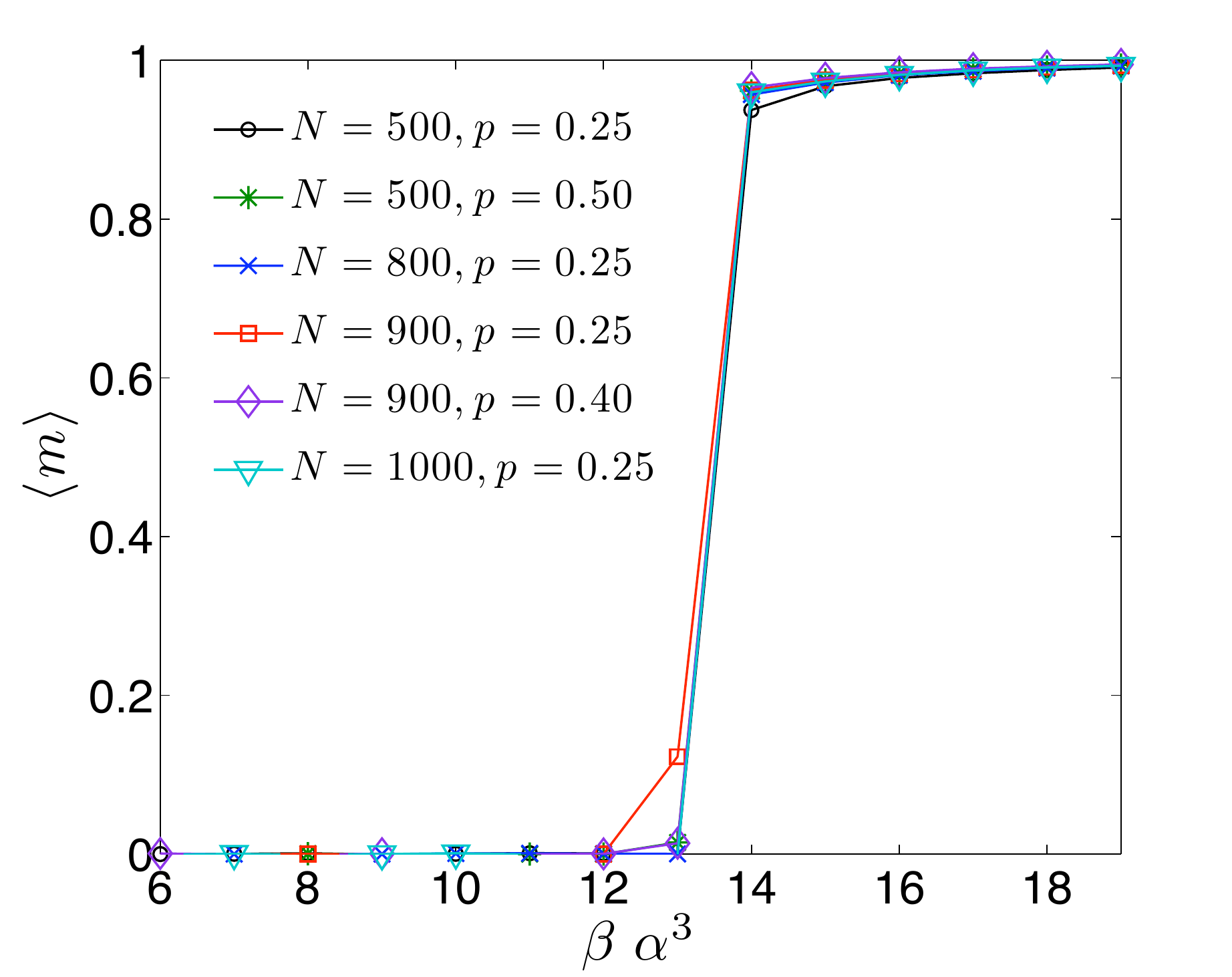}
\caption{\label{fig:p4}Magnetization for $4$-spin systems of different sizes and
different dilution as a function of $\beta \; \alpha^{p-1}$. The
collapse of all the curves provides a strong evidence for the
scaling of the temperature.} \ec
\end{figure}

In order to deepen the role of connectivity in the evolution of
the system we measure the macroscopic observable $\langle m \rangle$ and its (normalized) fluctuations $\langle m^2 \rangle -
\langle m \rangle^2$, studying their dependence on $T$ and on $\alpha$. Data for different choices of
size and dilution are shown in Figure \ref{fig:p3} for $p=3$ and in Figure \ref{fig:p4} for $p=4$.

The profile of the magnetization, with an abrupt jump, and the
correspondent peak found for its fluctuations confirm the
existence of a first order phase transition at a well defined
temperature $T_c$ whose value depends on the dilution $\alpha$.
More precisely, by properly normalizing the temperature in
agreement with analytical results, namely $\tilde\beta \equiv
\beta \; \alpha^{p-1}$ we found a very good collapse of all
the curves considered. Hence, we have agreement among analytic and
numerics concerning the scaling of the temperature as
$\alpha^{p-1}$. Moreover our data provide a very clear hint
suggesting that the critical temperature can be written as $T_c =
f(p) \alpha^{p-1}$, where $f(p)$ is a monotonic decreasing function of $p$.

\section{Conclusions}
In this paper we performed an analysis of the ferromagnetic
diluted $p$-spin model via cavity field technique and numerical
simulations. Several questions have been addressed, including an expression
for the free energy, the self-averaging families for the order
parameters and a study of the phase transition among a
paramagnetic and a ferromagnetic regime. Despite a
rigorous picture for the lacking of replica symmetry breaking in
diluted ferromagnet is still unavailable, we supported strong
evidence toward a full replica symmetric behavior in the whole
phase diagram. In particular, we showed the vanishing of criticality for $p>2$ and we found a proper scaling for the transition temperature as a function of the system dilution, namely $T_c \sim  \alpha^{p-1}$.

Further development should be two-fold: from one side the same
analysis is still to be performed on the X-OR-SAT model which
constitutes another element making up the class of models based on
binary agents with mean filed interaction. On the other side, the
whole mathematical architecture still suffers a not exhaustive
development; in fact the difference among even and odd $p$ model,
at least for large $p$, is thermodynamically almost irrelevant,
while the lacking of the gauge symmetry in the latter rules out
the method at this stage. Moreover, it is highlighted the need to
develop a Hamilton-Jacobi technique \cite{BG} in order to handle
this kind of problem to avoid the iteration procedure implied by
the cavity method.

\section{Appendix: Analytical proofs}

In this section the proofs of al the Theorems and the Proposition
$1$ are reported.

\bigskip

\textbf{Proof of Theorem \ref{primolevi}}
\newline
Bridging a system made of by $N+1$ spins with one made of by $N$
spins implies the definition of rescaled $\gamma, \alpha$
parameters, accordingly to \cite{ABC}\cite{BCC}
\begin{eqnarray}
\tilde{\gamma} &=& \gamma \frac{N}{N+1} \qquad
\stackrel{N\rightarrow\infty}{\longrightarrow} \quad \gamma \\
\tilde{\alpha} &=& \alpha \Big[\frac{N}{N+1}\Big]^{\frac{1}{p-1}}
\qquad \stackrel{N\rightarrow\infty}{\longrightarrow} \quad
\alpha.
\end{eqnarray}
We have, in distribution, the Hamiltonian of a system made of
$N+1$ particles writable as
\begin{eqnarray}\label{Hsplit}
H_{N+1}(\sigma,\gamma) &=& -\sum_{\nu=1}^{k_{\gamma (N+1)}}
\sigma_{i_{\nu}^1}\sigma_{i_{\nu}^2}...\sigma_{i_{\nu}^p}\,\sim\,
\\ \nonumber &-& \sum_{\nu=1}^{k_{\tilde{\gamma} N}}
\sigma_{i_{\nu}^1}\sigma_{i_{\nu}^2}...\sigma_{i_{\nu}^p} -
\sum_{\nu=1}^{k_{2\tilde{\gamma}}}
\sigma_{i_{\nu}^1}\sigma_{i_{\nu}^2}...\sigma_{i_{\nu}^{p-1}}\sigma_{N+1},
\end{eqnarray}
that we may rewrite as \be H_{N+1}(\sigma,\gamma) =
H_N(\sigma,\tilde{\gamma}) + \hat{H}_N(\sigma,2\tilde{\gamma}).
\ee
\medskip
\newline
Following the above decomposition, let us consider the partition
function of the same $N+1$ spin model and let us introduce the
gauge transformation $\sigma_{i} \rightarrow \sigma_i\sigma_{N+1}$
which is a symmetry of the Hamiltonian known as
\textit{spin-flip}.
\begin{eqnarray}
Z_{N+1}(\gamma,\beta) &\sim& \sum_{\{\sigma_{N+1}\}} e^{-\beta
H_{N}(\sigma,\tilde{\gamma}) -
\beta\hat{H}_{N}(\sigma,\tilde{\gamma})\sigma_{N+1}} = \label{T11}
\\ \nonumber &=& \sum_{\{\sigma_{N+1}\}}
e^{\beta H_{N}(\sigma,\tilde{\gamma}) +
\beta\sum_{\nu=1}^{k_{2\tilde{\gamma }}}
\sigma_{i_{\nu}^1}...\sigma_{i_{\nu}^{p-1}}\sigma_{N+1}} =
\label{T12} \\ \nonumber &=& 2 \sum_{\{\sigma_{N}\}}
e^{\beta\sum_{\nu=1}^{k_{\tilde{\gamma}N}}
\sigma_{i_{\nu}^1}...\sigma_{i_{\nu}^p} +
\beta\sum_{\nu=1}^{k_{2\tilde{\gamma }}} \sigma_{i_{\nu}^1}...\sigma_{i_{\nu}^{p-1}}} = \label{T13} \\
&=& 2 Z_N(\tilde{\gamma},\beta)\tilde{\omega}(e^{-\beta
\hat{H}_{N}}),\nonumber
\end{eqnarray}
where  the new Boltzmann state $\tilde{\omega}$, and its
replicated $\tilde{\Omega}$, are introduced as \begin{eqnarray}
\tilde{\omega}(g(\sigma)) &=&
\frac{\sum_{\{\sigma_{N}\}}g(\sigma)e^{-\beta
H_N(\tilde{\gamma},\sigma)}} {\sum_{\{\sigma_{N}\}}e^{-\beta
H_N(\tilde{\gamma},\sigma)}}, \\  \qquad \tilde{\Omega}(g(\sigma))
&=& \prod_i\tilde{\omega}^{(i)}(g(\sigma^{(i)})). \end{eqnarray}
\newline
To continue the proof we now take the logarithm of both sides of
the last expression in eq. (\ref{T11}), apply the expectation
$\textbf{E}$ and subtract the quantity $\textbf{E}[\ln
Z_{N+1}(\tilde{\gamma},\beta)]$. We obtain
\begin{eqnarray}\nonumber
&& \textbf{E}[\ln Z_{N+1}(\gamma,\beta)] - \textbf{E}[\ln
Z_{N+1}(\tilde{\gamma},\beta)]=  \\  && \ln2 - \textbf{E}[\ln
\frac{Z_{N+1}(\gamma,\beta)}{Z_N(\tilde{\gamma},\beta)}] +
\Psi_N(\tilde{\gamma},\beta,1),
\end{eqnarray}
The left hand side gives
\begin{eqnarray}
\textbf{E}[\ln Z_{N+1}(\gamma,\beta)] &-& \textbf{E}[\ln
Z_{N+1}(\tilde{\gamma},\beta)] = \\ \nonumber &=&
(\gamma - \tilde{\gamma})\frac{d}{d\gamma}\textbf{E}[\ln Z_{N+1}(\gamma,\beta)]|_{\gamma=\tilde{\gamma}} = \nonumber \\
&=& \gamma\frac{1}{N+1}\frac{d}{d\gamma}\textbf{E}[\ln Z_{N+1}(\gamma,\beta)]|_{\gamma=\tilde{\gamma}} = \nonumber \\
&=& \gamma \frac{d}{d\gamma}A_{N+1}(\gamma,\beta).
\end{eqnarray}

Considering the $\alpha$ dependence of $\gamma$, we have
$$
\partial_{\gamma} \propto \frac{1}{(p-1)\alpha^{p-2}}\partial_{\alpha} \quad
\Rightarrow \quad \gamma\frac{d}{d\gamma}A \propto
\frac{\alpha}{p-1}\frac{d}{d\alpha}A,
$$
where the symbol $\propto$ instead of $=$ reflects the
arbitrariness by which we include  the $p!$ term, multiplying
$\alpha$, inside the definition of $\gamma$, or directly in
$\alpha$.
\newline
Performing now the thermodynamic limit, we see that at the right
hand side we have \be \lim_{N \to \infty}\textbf{E}[\ln
\frac{Z_{N+1}(\alpha,\beta)}{Z_N(\tilde{\alpha},\beta)}]
\longrightarrow A(\alpha,\beta) \ee and the theorem is proved
$\Box$.

\bigskip

\textbf{Proofs of Theorems
\ref{ciccia},\ref{saturabili},\ref{saturi}}
\newline
In this sketch we are going to show how to get Theorem
(\ref{ciccia}) in some details; It automatically has as a
corollary Theorem (\ref{saturabili}) which ultimately gives, as a
simple consequence when applied on filled monomials,
Theorem(\ref{saturi}).
\newline
Let us assume for a generic overlap correlation function $Q$, of
$s$ replicas, the following representation
$$
Q = \prod_{a=1}^s\sum_{i_l^a}\prod_{l=1}^{n^a}\sigma_{i_l^a}^a
I(\{i_l^a \})
$$
where $a$ labels the replicas, the internal product takes into
account the spins (labeled by $l$) which contribute to the a-part
of the overlap $q_{a,a'}$ and runs to the number of time that the
replica $a$ appears in $Q$. The external product takes into
account all the contributions of the internal one and the $I$
factor fixes the constraints among different replicas in $Q$; so,
for example, $Q=q_{12}q_{23}$ can be decomposed in this form
noting that $s=3$, $n^1=1,n^2=2$,
$I=N^{-2}\delta_{i_1^1,i_1^3}\delta_{i_1^2,i_2^3}$, where the
$\delta$ functions fixes the links between replicas $1,3
\rightarrow q_{1,3}$ and $2,3 \rightarrow q_{2,3}$. The averaged
overlap correlation function is
$$
\langle Q \rangle_t = \mathbf{E}\sum_{i_l^a}I(\{i_l^a
\})\prod_{a=1}^s \omega_{t}(\prod_{l=1}^{n^a}\sigma_{i_l^a}^a).
$$
Now if $Q$ is a fillable polynomial, and we evaluate it at $t=1$,
let us decompose it, using the factorization of the $\omega$ state
on different replica, as
$$ \langle Q \rangle_t = \mathbf{E}\sum_{i_l^a,i_l^b}I(\{i_l^a \}, \{i_l^b \})\prod_{a=1}^u
\omega_a ( \prod_{l=1}^{n^a}\sigma_{i_l^a}^a) \prod_{b=u}^s
\omega_b ( \prod_{l=1}^{n^b}\sigma_{i_l^b}^b),
$$
where $u$ stands for the number of the unfilled replicas inside
the expression of $Q$. So we split the measure $\Omega$ into two
different subset $\omega_{a}$ and $\omega_{b}$: in this way the
replica belonging to the $b$ subset are always in even number,
while the ones in the $a$ subset are always odds. Applying the
gauge $\sigma_i^a \rightarrow \sigma_i^a\sigma_{N+1}^a, \forall i
\in (1,N)$ the even measure is unaffected by this transformation
$(\sigma_{N+1}^{2n} \equiv 1)$ while the odd measure takes a
$\sigma_{N+1}$ inside the Boltzmann measure. \begin{eqnarray} &&
\langle Q \rangle = \\ \nonumber && \sum_{i_l^a,i_l^b}I(\{i_l^a
\}, \{i_l^b \}) \prod_{a=1}^u \omega ( \sigma_{N+1}^a
\prod_{l=1}^{n^a}\sigma_{i_l^a}^a) \prod_{b=u}^s \omega (
\sigma_{N+1}^b\prod_{l=1}^{n^b}\sigma_{i_l^b}^b). \end{eqnarray}
At the end we can replace in the last expression the index $N+1$
of $\sigma_{N+1}$ by $k$ for any $k \neq \{ i_l^a \}$ and multiply
by one as $1=N^{-1}\sum_{k=0}^N$. Up to orders $O(1/N)$, which go
to zero in the thermodynamic limit, we have the proof.
\medskip
\newline
It is now immediate to understand that Theorem (\ref{ciccia}) on a
fillable overlap monomial has the effect of multiplying it by its
missing part to be filled (Theorem \ref{saturabili}), while it has
no effect if the overlap monomial is already filled (Theorem
\ref{saturi}). $\Box$

\bigskip

\textbf{Proof of Proposition \ref{stream}}
\newline
The proof works by direct calculation:
\begin{eqnarray}
&& \frac{\partial\langle F_s \rangle_{t,\tilde{\alpha}}}{\partial
t} = \\ \nonumber && \frac{\partial \textbf{E}}{\partial t}  \Big[
\frac{\sum_{\{\sigma\}}F_s
e^{\sum_{a=1}^s(\beta\sum_{\nu=1}^{k_{\tilde{\gamma} N}}
\sigma_{i_{\nu}^1}^a...\sigma_{i_{\nu}^p}^a +
\beta\sum_{\nu=1}^{k_{2\tilde{\gamma}t}}
\sigma_{i_{\nu}^1}^a...\sigma_{i_{\nu}^{p-1}}^a)}}
{\sum_{\{\sigma\}}
e^{\sum_{a=1}^s(\beta\sum_{\nu=1}^{k_{\tilde{\gamma} N}}
\sigma_{i_{\nu}^1}^a...\sigma_{i_{\nu}^p}^a + \beta
\sum_{\nu=1}^{k_{2\tilde{\gamma}t}}
\sigma_{i_{\nu}^1}^a...\sigma_{i_{\nu}^{p-1}}^a)}}\Big] =
\\ \nonumber && 2\tilde{\alpha}^{p-1} \textbf{E}\Big[
\frac{\tilde{\Omega}_t (F_s
e^{\sum_{a=1}^s(\beta\sigma_{i_{0}^1}^a...\sigma_{i_{0}^{p-1}}^a)})}
{\tilde{\Omega}_t(e^{\sum_{a=1}^s(\beta\sigma_{i_{0}^1}^a...\sigma_{i_{0}^{p-1}}^a)})}\Big]
- 2\tilde{\alpha}^{p-1}\langle F_s \rangle_{t,\tilde{\alpha}} = \\
\nonumber &&  2\tilde{\alpha} \textbf{E}\Big[
\frac{\tilde{\Omega}_t (F_s \Pi_{a=1}^{s}(\cosh\beta +
\sigma_{i_{0}^1}^a...\sigma_{i_{0}^{p-1}}^a\sinh\beta))}
{\tilde{\Omega}_t (\Pi_{a=1}^{s}(\cosh\beta +
\sigma_{i_{0}^1}^a...\sigma_{i_{0}^{p-1}}^a\sinh\beta))}\Big]- \\
\nonumber &&
2\tilde{\alpha}^{p-1}\langle F_s \rangle_{t,\tilde{\alpha}} = \\
\nonumber &&   2\tilde{\alpha}^{p-1} (\textbf{E}\Big[
\frac{\tilde{\Omega}_t (F_s \Pi_{a=1}^{s}(1 +
\sigma_{i_{0}^1}^a...\sigma_{i_{0}^{p-1}}^a\theta))} {(1 +
\tilde{\omega}_t(\sigma_{i_{0}^1}^a...\sigma_{i_{0}^{p-1}}^a)\theta)^s}\Big]
- \langle F_s \rangle_{t,\tilde{\alpha}}),
\end{eqnarray}
Now noting that
\begin{eqnarray} \nonumber
\Pi_{a=1}^{s}(1 &+&
\sigma_{i_{0}^1}^a...\sigma_{i_{0}^{p-1}}^a\theta) = 1 +
\sum_{a=1}^{s}\sigma_{i_{0}^1}^a...\sigma_{i_{0}^{p-1}}^a\theta
\\ \nonumber &+& \sum_{a<b}^{1,s}\sigma_{i_{0}^1}^a...\sigma_{i_{0}^{p-1}}^a
\sigma_{i_{0}^1}^b...\sigma_{i_{0}^{p-1}}^b\theta^2
+ ...\nonumber \\
\frac{1}{(1 + \tilde{\omega}_t \theta)^s} &=& 1 -
s\tilde{\omega}_t \theta + \frac{s(s+1)}{2!}\tilde{\omega}_t^2
\theta^2 + ... \nonumber
\end{eqnarray}
\medskip
we obtain
\begin{eqnarray}
\frac{\partial\langle F_s \rangle_{t,\tilde{\alpha}}}{\partial t}
&=& 2\tilde{\alpha}^{p-1} \Big(\textbf{E}\Big[ \tilde{\Omega}_t
\Big(F_s(1 +
\sum_{a=1}^{s}\sigma_{i_{0}^1}^a...\sigma_{i_{0}^{p-1}}^a\theta +
\\ \nonumber &+& \sum_{a<b}^{1,s}\sigma_{i_{0}^1}^a...\sigma_{i_{0}^{p-1}}^a
\sigma_{i_{0}^1}^b...\sigma_{i_{0}^{p-1}}^b\theta^2
+ ...)\Big) \times \nonumber \\
&\times& \Big(1 - s\tilde{\omega}_t \theta +
\frac{s(s+1)}{2!}\tilde{\omega}_t^2 \theta^2 + ...\Big)\Big] -
\langle F_s \rangle_{t,\tilde{\alpha}}\Big), \nonumber
\end{eqnarray}
from which our thesis follows. $\Box$

\section*{Acknowledgment}

The authors are grateful to Francesco Guerra, Pierluigi Contucci
and Raffaella Burioni for interesting discussions.

\end{document}